\documentclass[doublecol]{epl2}

\usepackage{ifthen}
\usepackage{ifpdf}

\usepackage{latexsym}
\usepackage{amssymb} 
\usepackage{bm}

\ifpdf
\usepackage{graphicx}
\usepackage{epstopdf}
\else
\usepackage{graphicx}
\usepackage{epsfig}
\fi

\newcommand{\eexp}{\,\mbox{e}^}

\newcommand{\mass}{\mathsf{M}} 

\newcommand{\tbox}[1]{\mbox{\tiny #1}}

\newcommand{\be}[1]{\begin{eqnarray}\ifthenelse{#1=-1}{\nonumber}{\ifthenelse{
#1=0}{}{\label{e#1}}}}
\newcommand{\ee}{\end{eqnarray}} 

\newcommand{\hide}[1]{}

\title{Semilinear response for the heating rate \\ of cold atoms in vibrating traps}
\shorttitle{Semilinear response}

\author{Alexander Stotland$^1$, Doron Cohen$^1$ and Nir Davidson$^2$}

\institute{
$^1$Department of Physics, Ben-Gurion University, Beer-Sheva, 84005, Israel \\
$^2$Department of Physics of Complex Systems, Weizmann Institute of Science, Rehovot, 76100 Israel
}

\pacs{03.65.-w}{Quantum mechanics}

\abstract{
The calculation of the heating rate of cold atoms in vibrating traps 
requires a theory that goes beyond the Kubo linear response formulation. 
If a strong ``quantum chaos'' assumption does not hold, the analysis of
transitions shows similarities with a percolation problem in 
energy space. We show how the texture and the sparsity of the perturbation matrix, 
as determined by the geometry of the system, dictate the result. 
An improved sparse random matrix model is introduced: it captures the essential 
ingredients of the problem, and leads to a generalized variable range hopping
picture.
}

\begin{document} 
\maketitle


The rate of energy absorption by particles that are confined
by vibrating walls was of interest in past studies
of nuclear friction \cite{wall1,wall2,frc}, 
where it leads to the damping of the wall motion. 
More recently it has become of interest
in the context of cold atoms physics. 
In a series of experiments \cite{nir1,nir2,nir3}  
with ``atom-optics billiards"
some surprising predictions \cite{wls} based 
on linear response theory (LRT) have been verified.

In the present study we consider the case
where the billiard is fully chaotic \cite{a} 
but with nearly integrable shape (Fig.\ref{fig:model}).
We explain that in such circumstances LRT is {\em not} 
applicable (unless the driving is extremely weak 
such that relaxation dominates).  Rather, the   
analysis that is relevant to the typical experimental 
conditions should go beyond LRT, and involve 
a ``resistor network" picture of transitions
in energy space, somewhat similar to a percolation problem. 
Consequently we predict that the rate of energy 
absorption would be suppressed by orders of magnitude, 
and provide some analytical estimates which 
are supported by a numerical calculation.

We assume that an experimentalist has control over the position ($R$) 
of a wall element that confines the motion of cold atoms in an optical trap. 
We consider below the effect of low frequency noisy (non periodic) driving. 
This means that $R$ is not strictly constant in time, either because of
drifts \cite{drifts} that cannot be eliminated in realistic circumstances, 
or else deliberately as a way to probe the dynamics of the atoms inside the trap \cite{shake}.
We assume the usual Markovian picture of FGR transitions
between energy levels, which is applicable in typical circumstances  
(see e.g. \cite{louisell}).   
These transitions lead to diffusion in the energy space.
If the atomic cloud is characterized by a temperature~$T$,
then the diffusion in energy would lead to heating
with the rate $\dot{E}=D/T$ \cite{bb}
and hence to an increase in the temperature of the cloud.

Naively one expects to observe an LRT behavior. 
That means to have $D \propto [\mbox{RMS}(\dot{R})]^2$, 
and more specifically to have a linear relation 
between the diffusion coefficient and the power 
spectrum of the driving:
\be{1324}
D \ \ \equiv \ \ G \times \mbox{RMS}(\dot{R})^2  
\ \ = \ \ \int_{0}^{\infty} \tilde{C}(\omega) \tilde{S}(\omega) d\omega
\ee
Here $\tilde{S}(\omega)$ is the power spectrum of $\dot{R}$, 
and $\tilde{C}(\omega)$ is related to the susceptibility
of the system. From the experimentalist's point of view 
the second equality in Eq.(\ref{e1324}) 
can be regarded as providing a practical definition 
for $\tilde{C}(\omega)$, if the response is indeed linear.

We shall explain in this paper that the applicability 
of LRT in our problem is very limited, namely 
LRT would lead to wrong predictions 
in {\em typical} experimental circumstances.     
Rather we are going to use a more refined theory, 
which we call semi-linear response theory (SLRT) \cite{slr1,slr2},  
in order to determine~$D$. 
The theory is called SLRT because on the one hand  
the power spectrum ${\tilde{S}(\omega) \mapsto \lambda \tilde{S}(\omega)}$ 
leads to ${D \mapsto \lambda D}$, but on the other 
hand ${\tilde{S}(\omega) \mapsto \tilde{S}_1(\omega)+\tilde{S}_2(\omega)}$ 
does not lead to ${D \mapsto D_1 + D_2}$. 
This semi-linearity can be tested in an experiment 
in order to distinguish it from linear response.    
Accordingly, in SLRT the spectral function $\tilde{C}(\omega)$ 
of Eq.(\ref{e1324}) becomes ill defined, 
while the coefficient $G$ is still physically {\em meaningful}, 
and can be measured in an actual experiment.

If we assume small driving amplitude the Hamiltonian matrix 
can be written as ${\mathcal{H} = \{E_n\} + f(t) \{V_{nm}\}}$, 
where 
\be{0}
V_{nm} \ \ = \ \ \Big\langle n\Big|\frac{d\mathcal{H}}{dR}\Big|m\Big\rangle
\ee
is the perturbation matrix.  
More than 50 years ago Wigner had proposed  
to regard the perturbation matrix of a complex system  
as a random matrix (RMT) whose elements are taken from 
a Gaussian distribution. Later Bohigas had conjectured 
that the same philosophy applies to quantized chaotic systems.
For such matrices the validity of LRT can be established 
on the basis of the FGR picture, 
and the expression for $G$ is the Kubo formula 
$G_{\tbox{LRT}} = \pi \varrho_{\tbox{E}} \langle\langle |V_{nm}|^2 \rangle\rangle_{a}$
where ${\langle\langle x \rangle\rangle_a=\langle x \rangle}$
is the algebraic average over the near diagonal matrix elements \cite{b},
and $\varrho_{\tbox{E}}$ is the density of states (DOS).
In contrast to that, 
using the Pauli master equation \cite{louisell} 
with FGR transition rates between levels, 
the SLRT analysis leads to
\be{1}
G_{\tbox{SLRT}}  \ \ = \ \ \pi \varrho_{\tbox{E}}
\,\langle\langle |V_{nm}|^2 \rangle\rangle
\ee
where the ``average" $\langle\langle x \rangle\rangle$ 
is defined as in Ref.\cite{slr1,slr2}   
via a {\em resistor-network calculation} \cite{miller}.
(For mathematical details see ``the SLRT calculation" paragraph below).

Within the RMT framework an element $x$ of $|V_{nm}|^2$ 
is regarded as a random variable, and the histogram 
of all $x$~values is used in order to define an 
appropriate ensemble. For the sake of later discussion 
we define, besides the algebraic average ${\langle\langle x \rangle\rangle_a}$
also the harmonic average as ${\langle\langle x \rangle\rangle_h=[\langle1/x\rangle]^{-1}}$ 
and the geometric average as ${\langle\langle x \rangle\rangle_g=\exp[\langle\ln x\rangle]}$. 
The result of the resistor network calculation is 
labeled as $\langle\langle x \rangle\rangle$ (without subscript).

Our interest is in the circumstances where the 
strong ``quantum chaos" assumption of Wigner fails.
This would be the case if the distribution of $x$
is wide in log scale. 
If $x$ has (say) a log-normal distribution, 
then it means that the typical value of $x$ is much smaller compared 
with the algebraic average. 
This means that the perturbation matrix $V_{nm}$ is 
effectively sparse ({\em a lot of vanishingly small elements)}.
We can characterize the sparsity by the 
parameter ${q = \langle\langle x \rangle\rangle_g/\langle\langle x \rangle\rangle_a}$. 
We are going to explain that for typical experimental 
conditions we might encounter sparse matrices for which ${q\ll 1}$. 
Then the energy spreading process is similar to a percolation in energy space, 
and the SLRT formula Eq.(\ref{e1}) replaces the Kubo formula.

\section{Outline} 
In what follow we present our model system, 
analyze it within the framework of SLRT, 
and then introduce an RMT model with log-normal 
distributed elements, that captures the essential
ingredients of the problem. 
We show that a generalized resistor network 
analysis for the transitions in energy space  
leads to a generalized Variable Range Hopping (VRH) picture 
(the standard VRH picture has been introduced by Mott in \cite{mott} 
and later refined by \cite{ambeg} using 
the resistor network perspective of \cite{miller}).
Our RMT based analytical estimates are verified 
against numerical calculation.
Finally we discuss the experimental aspect, 
and in particular define the physical circumstances 
in which SLRT rather than LRT applies. 
These two theories give results that can differ 
by orders of magnitude.

\section{Modeling}
\begin{figure}[h!t]
\centering
\includegraphics[clip,width=0.75\hsize]{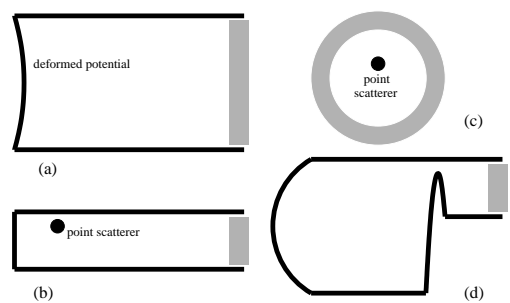}

\caption{Model systems: The atoms are held by a potential 
that may consist of static walls (solid lines),  
a vibrating wall (shaded lines), and bumps (thick points). 
The numerics has been done for (b) with a Gaussian bump.
We work with two different aspect ratios. 
For the aspect ratio $AS=20$ we take $L_x = 200$ and $L_y = 10$. 
For the aspect ratio $AS=1$ we take $L_x = 40$ and $L_y = 40$. 
The position of the Gaussian bump was randomly chosen within 
the region ${ [0.4,0.6]L_x \times [0.4,0.6]L_y}$. 
The width of the Gaussian is $\sigma_x = \sigma_y = \sigma$.  
We have assumed noisy driving with $\omega_c = 7\Delta$, 
where $\Delta=1/\varrho_{\tbox{E}}$ is the mean level spacing, 
and the units were such that $\mass=1$.}

\label{fig:model}
\end{figure}

Consider a strictly rectangular billiard whose 
eigenstates are labeled by ${\bm{n}=(n_x, n_y)}$.   
The perturbation due to the movement of 
the `vertical' wall does not couple states 
that have different mode index $n_y$. 
Due to this selection rule 
the perturbation matrix is sparse.  
If we deform slightly the potential (Fig.\ref{fig:model}a),
or introduce a bump (Fig.\ref{fig:model}b), 
then states with different mode index are mixed. 
Consequently the numerous zero elements become 
finite but still very tiny in magnitude, which means a very wide 
size distribution featuring a small fraction of large elements.
Similar considerations apply for 
the circular cavity of Fig.\ref{fig:model}c, 
where an off-center scatterer couples radial and angular motion,   
and which is more suitable for a real experiment 
(but less convenient for numerical analysis).

\begin{figure}[h!t]
\centering

\setlength{\unitlength}{1mm}
\noindent
\begin{picture}(85,65)
\put(0,0){\includegraphics[clip,width=85mm]{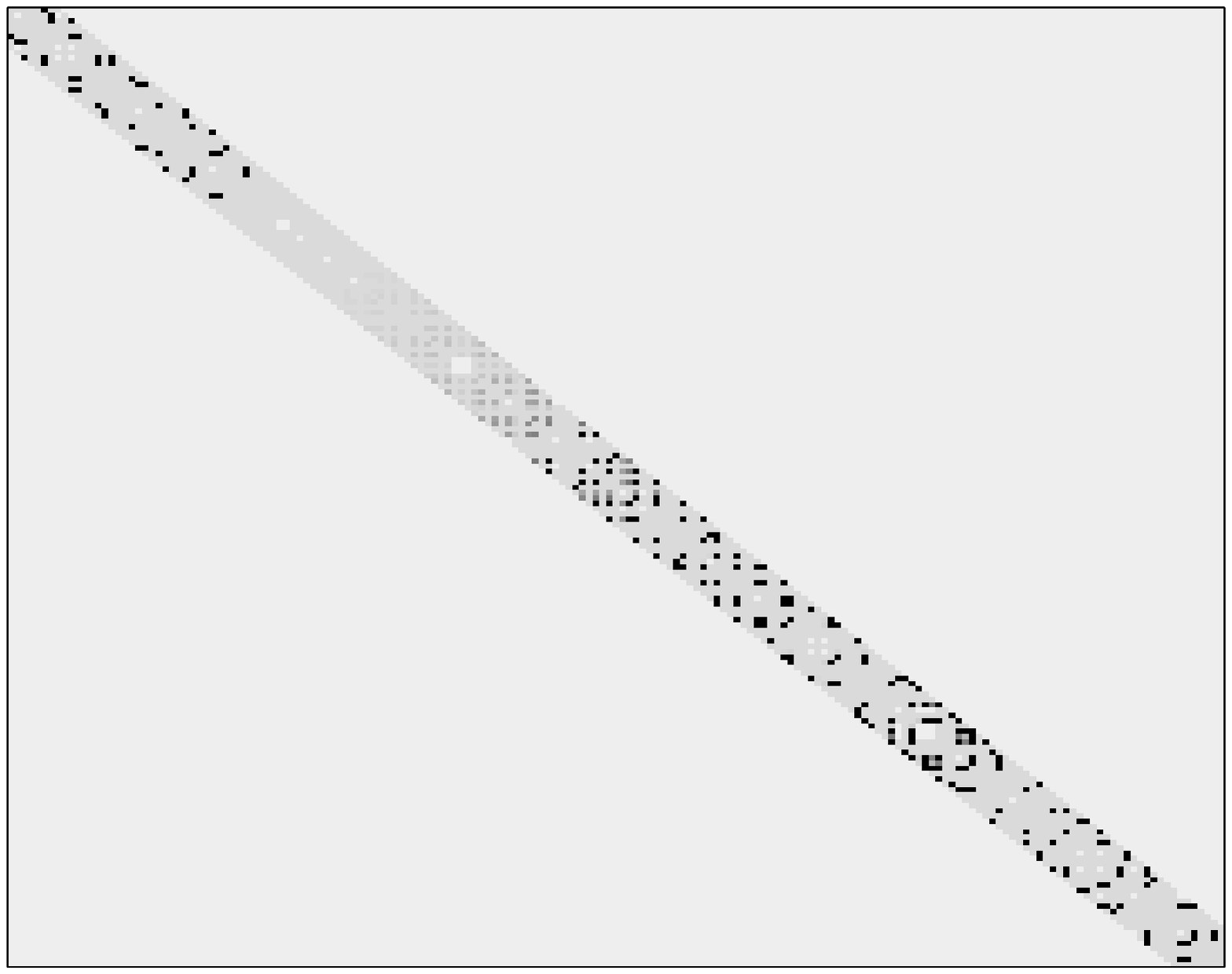}}
\put(0.5,0){\includegraphics[clip,width=40mm]{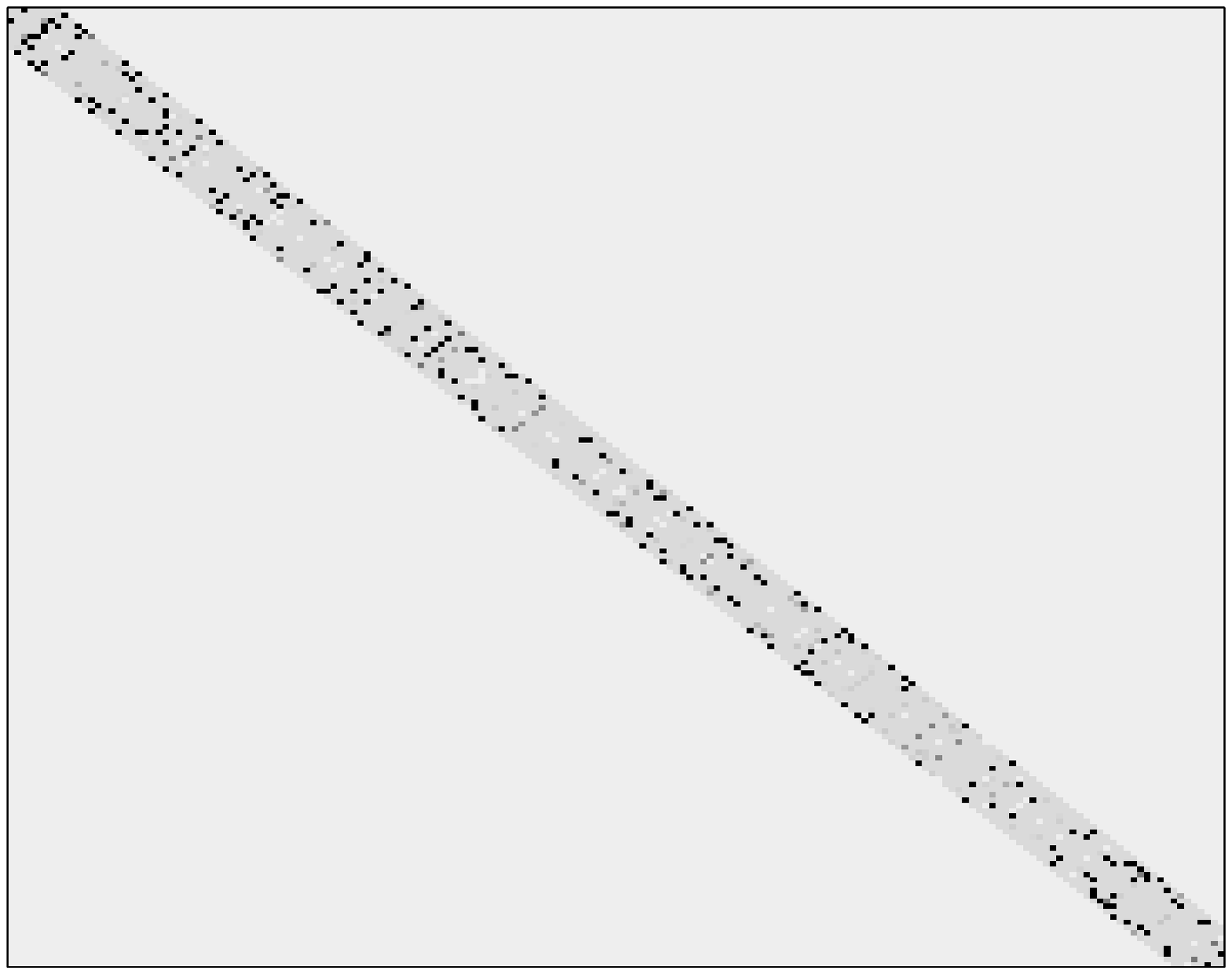}}
\put(44.5,35){\includegraphics[clip,width=40mm]{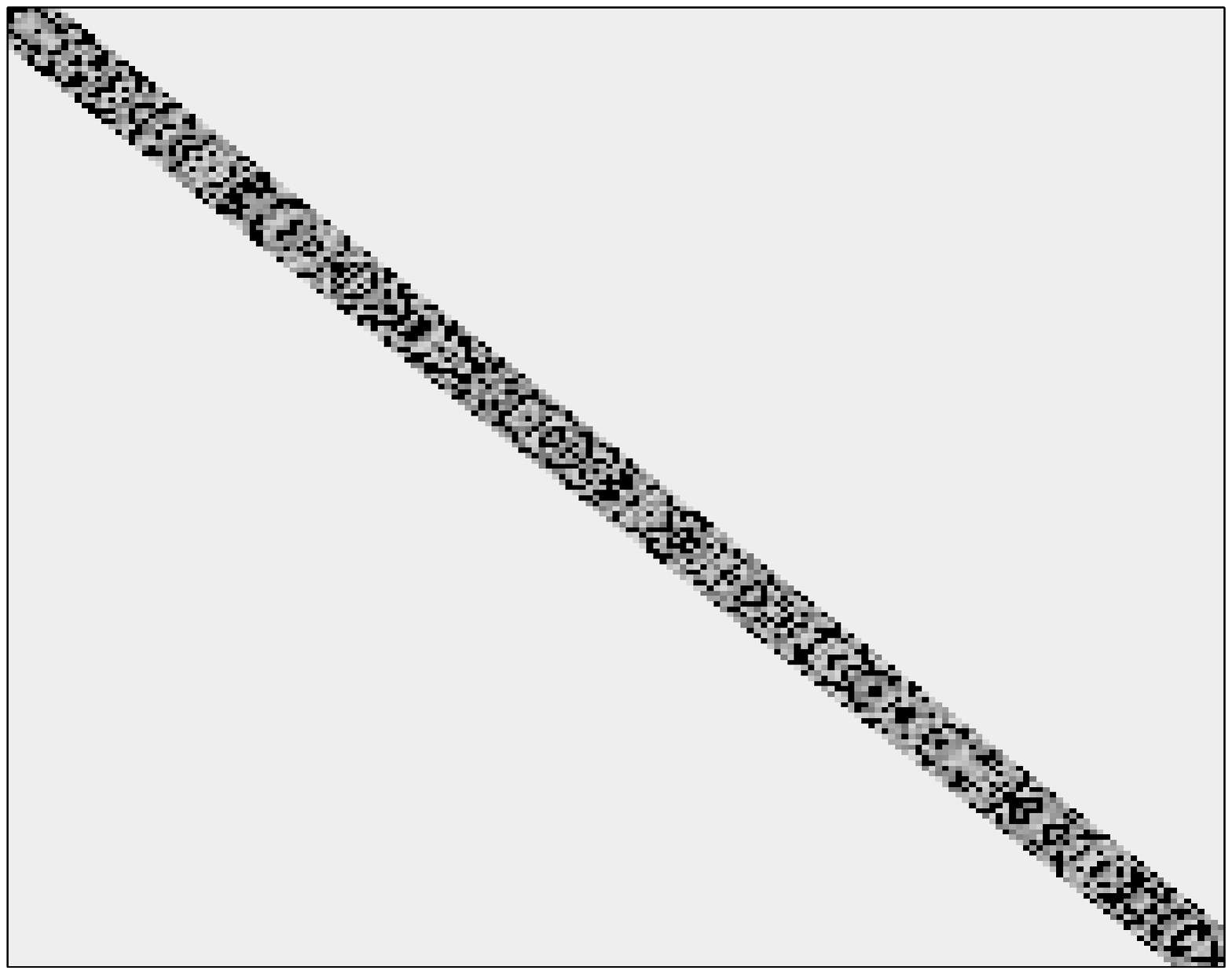}}
\end{picture}

\caption{
Image  of the perturbation matrix $|V_{nm}|^2$ due 
to a wall displacement of a rectangular-like cavity 
that has an aspect ratio ${AS=20}$. The potential floor 
is deformed due to the presence of a ${\sigma{=}0}$ scatterer
with ${u = 10^{-4}}$ (see text). The matrix is both 
sparse and textured. 
{\em Lower inset:} untextured matrix - the elements along each 
diagonal are randomly permuted. {\em Upper inset:} 
non-sparse matrix with the same band profile - 
each element is generated independently from a normal distribution.}

\label{fig:image}
\end{figure}

Typically the perturbation matrix 
is not only {\em sparse} but also {\em textured}. 
This means (see Fig.\ref{fig:image}) 
that there are stripes where the matrix elements 
are larger, and bottlenecks where they are all small.
The emergence of texture (i.e. non-random  
arrangement of the sparse large elements along the diagonals)  
is most obvious if we consider the geometry of Fig.\ref{fig:model}d, 
where we have a divided cavity with a small 
weakly connected chamber where the driving is applied. 
If the chamber were disconnected, then only chamber states 
with energies $E_r$ would be coupled by the driving. 
But due to the connecting corridor there is mixing 
of bulk states with chamber states within energy stripes 
around $E_r$. The coupling between two cavity 
states $E_n$ and $E_m$ is very small outside of 
the $E_r$ stripes. Consequently the near diagonal  
elements of $V_{nm}$ have wide variation, and hence 
a wide $\log(x)$ distribution. \hide{(though their 
arrangement is not random).}

Coming back to the geometries of Fig.\ref{fig:model}abc, 
it is somewhat important in the analysis 
to distinguish between {\em smooth} deformation 
that couples only nearby modes, 
and {\em diffractive} deformation 
that mix all the modes simultaneously: 
Recalling that different modes have different DOS,  
and that low-DOS modes are sparse within the high-DOS modes, 
we expect a more prominent manifestation of the texture  
in the case of a smooth deformation 
of a cavity that has a large aspect ratio.
We later confirm this expectation in the numerical analysis.

\section{The SLRT calculation}
As in the standard derivation of the Kubo formula, 
also within the framework of SLRT \cite{slr1,slr2}, 
the leading mechanism for absorption is assumed 
to be FGR transitions.  
The FGR transition rate is proportional to the 
squared matrix elements $|V_{nm}|^2$, 
and to the power spectrum at the frequency ${\omega=E_n{-}E_m}$. 
It is convenient to define the normalized spectral 
function $\tilde{F}(\omega)$, such that 
\be{0}
\tilde{S}(\omega) \ \ \equiv \ \ \mbox{RMS}(\dot{R})^2 \times \tilde{F}(\omega)
\ee
Contrary to the naive expectation the theory does not lead 
to the Kubo formula. This is because the rate of absorption 
depends crucially on the possibility to make {\em connected} 
sequences of transitions. It is implied that both the texture 
and the sparsity of the $|V_{nm}|^2$ matrix play a major role 
in the calculation of~$G$.
Consequently SLRT leads to Eq.(\ref{e1}), 
where ${\langle\langle...\rangle\rangle}$ 
is defined using a resistor network calculation. 
Namely, the energy levels are regarded 
as the {\em nodes} of a resistor network, 
and the FGR transition rates as the {\em bonds} 
that connect different nodes.
Following \cite{slr2} the inverse 
resistance of a bond is defined as   
\be{2}
\mathsf{g}_{nm}  \ \ \equiv\ \ 
2\varrho_{\tbox{E}}^{-3} \ 
\frac{|V_{nm}|^2}{(E_n{-}E_m)^2} \   
\tilde{F}(E_m{-}E_n)
\ee
and ${\langle\langle|V_{nm}|^2\rangle\rangle}$ 
is defined as the inverse resistivity of the network. 
It is a simple exercise to verify  
that if all the matrix elements are the same, say  ${|V_{nm}|^2 = c}$, 
then ${\langle\langle|V_{nm}|^2\rangle\rangle = c}$ too. 
But if the matrix is sparse or textured then typically 
\be{3489}
\langle\langle|V_{nm}|^2\rangle\rangle_{\tbox{h}}  
\ll \langle\langle|V_{nm}|^2\rangle\rangle  
\ll \langle\langle|V_{nm}|^2\rangle\rangle_{\tbox{a}}
\ee
In the case of sparse matrices this is a mathematically 
strict inequality, and we can use a generalized VRH scheme 
which we describe below in order to get 
an estimate for $\langle\langle x \rangle\rangle$. 
If the {\em element-size} distribution of $\log(x)$ is not too stretched 
a reasonable approximation is  
${\langle\langle x \rangle\rangle \approx \langle\langle x \rangle\rangle_{\tbox{g}}}$, 
simply because the geometric mean is the {\em typical} (median) 
value for the size of the elements. 
However, if $|V_{nm}|^2$ has either a very stretched element-size distribution,
or if it has texture, then our VRH analysis below show that the geometric 
average becomes merely an improved {\em lower bound} for the actual result.

\section{Analysis}
We consider a particle of mass $\mass$ in a two dimensional 
box of length~$L_x$ and width~$L_y$, 
such that ${0<x<L_x}$ and ${0<y<L_y}$. See Fig.\ref{fig:model}b.
With the driving the length of the box becomes ${R=L_x+f(t)}$. 
The Hamiltonian is 
\be{0}
\mathcal{H} =  \mbox{diag}\{ E_{\bm{n}} \} + u \{ U_{\bm{n}\bm{m}} \} 
+ f(t) \{V_{\bm{n}\bm{m}}\}
\ee
where $\bm{n}=(n_x,n_y)$ is a composite index that labels the 
energy levels $E_{\bm{n}}$ of a particle in a rectangular box 
of size ${L_x \times L_y}$. 
The deformation is described by a normalized Gaussian potential~$U(x,y)$  
of width $(\sigma_x, \sigma_y)$ positioned at the central region of the box.  
Its matrix elements are $U_{\bm{n}\bm{m}}$, and  
it is multiplied in the Hamiltonian by a parameter~$u$ 
which signifies the strength of the deformation.
Note that the limit ${\sigma\rightarrow0}$ is well defined 
and corresponds to an ``s-scatterer".   
The perturbation matrix due to the $f(t)$ displacement 
of the wall is    
\be{223}
V_{\bm{n}\bm{m}} \ \ = \ \ -\delta_{n_y, m_y} \times \frac{\pi^2}{\mass L_x^3}
n_x m_x
\ee
The power spectrum of $\dot{f}$ is assumed to be constant within 
the frequency range $|\omega|<\omega_c$ and zero otherwise.
This means that ${\tilde{F}(\omega)=1}$ up to this cutoff frequency. 
We have also considered (not presented) an exponential 
line shape ${\tilde{F}(\omega)=\exp(-|\omega/\omega_c|)}$, 
leading to qualitatively similar results.
After diagonalization of ${\{E_{\bm{n}}\}+u\{U_{\bm{n}\bm{m}}\}}$ 
the Hamiltonian takes the form
\be{0}
\mathcal{H} = \mbox{diag}\{ E_n \}  + f(t) \{V_{nm}\}
\ee
where $n$ (not bold) is a running index that counts 
the energies in ascending order. The DOS  
remains essentially the same as for ${u=0}$, namely, 
\be{0}
\varrho_{\tbox{E}}  \ \ = \ \ \frac{1}{2\pi} \mass L_x L_y
\ee
The perturbation matrix $|V_{nm}|^2$ is sparse and textured (see Fig.\ref{fig:image}). 
First we discuss the sparsity, and the effect of the texture will be addressed later on.

\begin{figure}[h!t]
\centering

\includegraphics[clip,width=0.7\hsize]{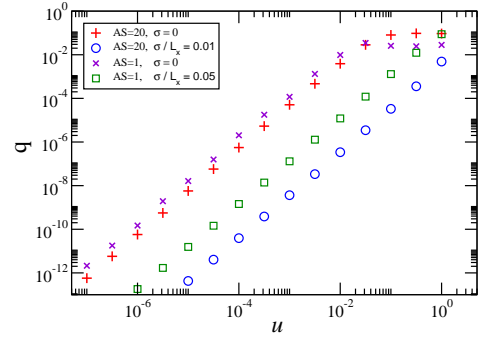}

\caption{
The sparsity parameter~$q$ is plotted versus the 
strength~$u$ of the deformation potential for 
cavities with aspect ratios ${AS=1}$ and ${AS=20}$. 
We see that for large aspect ratio $q$~has 
some sensitivity to~$\sigma$. 
As explained in the text $G_{\tbox{SLRT}}/G_{\tbox{LRT}}$ 
is correlated with~$q$, but for large aspect ratio 
it is even more sensitive to~$\sigma$ 
due to the emergence of textures whose presence 
is {\em not} reflected by the value of~$q$.}
\label{fig:qvsu}
\end{figure}

Considering first zero deformation ($u=0$) it follows  
from Eq.(\ref{e223}) that the non-zero elements
of the perturbation matrix 
are ${|V_{nm}|^2 \approx |\mass v_{\tbox{E}}^2 / L_x|^2}$, 
where ${v_{\tbox{E}}=\sqrt{2E/\mass}}$.
The algebraic average of the near diagonal elements equals 
this value (of the large size elements) 
multiplied by their percentage~$p_0$. 
To evaluate~$p_0$ let us consider an energy window~$dE$. 
The number of near-diagonal elements $V_{nm}$ 
within the stripe ${|E_{n_x,n_y}{-}E_{m_x,m_y}|<d\varepsilon}$ 
is ${\varrho_{\tbox{E}}^2 dE d\varepsilon}$. 
It is a straightforward exercise to find out that 
the the number of non-zero elements (i.e. with ${n_y{=}m_y}$) 
is the same number multiplied by ${p_0=[2\pi\mass v_{\tbox{E}} L_y]^{-1}}$. 
Consequently  
\be{21}
\langle\langle |V_{nm}|^2 \rangle\rangle_a 
\approx 
\left[\frac{1}{2\pi\mass v_{\tbox{E}} L_y}\right]
\left|\frac{\mass v_{\tbox{E}}^2}{L_x}\right|^2
= \frac{\mass v_{\tbox{E}}^3}{2\pi L_y L_x^2}
\ee
Somewhat surprisingly this result turns out to be the same 
(disregarding an order unity numerical prefactor)
as for a strongly chaotic cavity (see Eq.(I3) of Ref.\cite{frc}), 
as if there is no sparsity issue. This implies 
that irrespective of the deformation~$u$, 
the LRT Kubo result is identical to the 2D version 
of the wall formula (see Sec.7 of Ref.\cite{frc}): 
\be{23}
G_{\tbox{LRT}} \ \ = \ \  \frac{4}{3\pi} \frac{\mass^2 v_{\tbox{E}}^3}{L_x}
\ee
Our interest below is not in $G_{\tbox{LRT}}$ 
but in $G_{\tbox{SLRT}}$, 
which can differ by many orders of magnitudes.
For sufficiently small~$u$ the large size matrix elements
are not affected, and therefore the algebraic 
average stays the same. But in the SLRT calculation 
we care about the small size matrix elements, 
that are zero if ${u=0}$. Due to the first-order mixing 
of the levels, the typical overlap  ${|\langle \bm{m} | n \rangle|}$  
between perturbed and unperturbed states 
is ${|uU_{\bf{n}\bf{m}}/(E_{\bm{n}}{-}E_{\bm{m}})|}$. 
The typical size of a small $V_{nm}$ element 
is the multiplication of this overlap (evaluated for 
nearby levels) by the size 
of a {\em non-zero}  $V_{\bm{n}\bm{m}}$ element.  
Consequently the small size matrix elements 
are proportional to~$u^2$. 
The geometric average simply equals their typical size, leading to 
\be{22}
\langle\langle |V_{nm}|^2 \rangle\rangle_g \approx  
\left(\frac{\mass^2 v_{\tbox{E}}^2}{2\pi L_x}\right)^2 
\eexp{-2\mass^2 v_{\tbox{E}}^2 (\sigma_x^2 + \sigma_y^2)}
\ u^2
\ee
Motivated by the discussion below Eq.(\ref{e3489})  
a crude estimate for the SLRT result is 
${ G_{\tbox{SLRT}} \approx q \times G_{\tbox{LRT}}}$, 
where for small deformation  
\be{24}
q \ \ = \ \ 
\frac
{\langle\langle |V_{nm}|^2 \rangle\rangle_g}
{\langle\langle |V_{nm}|^2 \rangle\rangle_a} 
\ \ \propto \ \ u^2 
\ \ \ \ \ \ \ 
\mbox{see Eqs.(\ref{e21},\ref{e22})}
\ee
It follows from the above (and see Fig.\ref{fig:qvsu}) that 
for small deformations $q\ll1$, and consequently 
we expect ${G_{\tbox{SLRT}} \ll G_{\tbox{LRT}}}$.
This should be contrasted with the case of 
strongly deformed box for which all the elements 
are of the same order of magnitude and $q$ becomes 
of order unity. Our next task is to further 
improve the SLRT estimate using a proper 
resistor network calculation \cite{cc}.

\section{RMT modeling}
\begin{figure}[h!t]
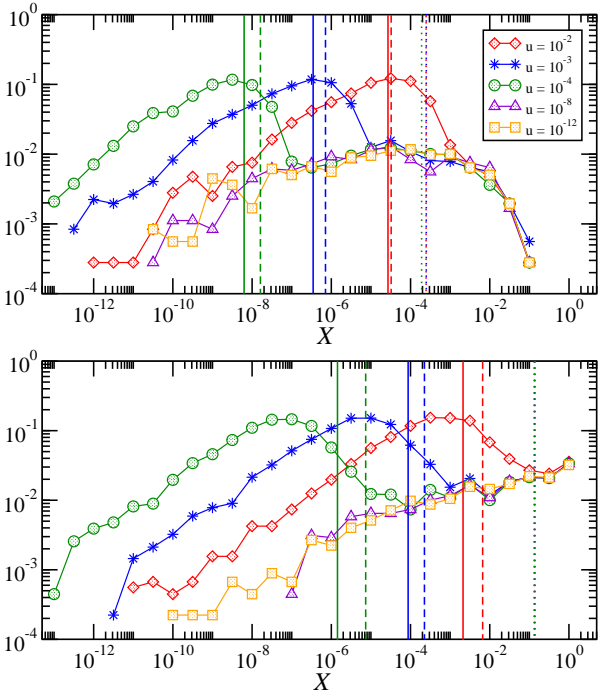

\centering

\includegraphics[clip,width=0.9\hsize]{hist_delta_ratio1}

\includegraphics[clip,width=0.9\hsize]{hist_delta_ratio20}

\caption{
Histograms of matrix elements for different values of $u$ 
for ${AS=1}$ (upper) and ${AS=20}$ (lower). 
Here we assume a ${\sigma=0}$ scatterer. 
The vertical lines for ${u=10^{-2},10^{-3},10^{-4}}$  
indicate the $\langle\langle x \rangle\rangle$  
obtained from the LRT algebraic average 
(3~dotted lines that are barely resolved),  
from the SLRT resistor network calculation (solid lines),  
and from the untextured calculation (dashed lines). 
The geometric mean approximately coincides 
with the peaks, and underestimates the SLRT value 
for the larger $AS$ where the sparsity is much larger.  
}

\label{fig:hist_delta_ratio_comp}
\end{figure}

The  $|V_{nm}|^2$ matrix looks like a random matrix with some 
distribution for the size of the elements (see Fig.\ref{fig:hist_delta_ratio_comp}). 
It might also possess some non-trivial texture which we ignore  
within the RMT framework. The RMT perspective allows 
us to derive a quantitative theory for~$G$ 
using a generalized VRH estimate. 
Let us demonstrate the procedure in the case 
of an homogeneous (neither banded nor textured)  
random matrix with log-normal distributed elements.
The mean and the variance of $\ln(x)$ are trivially related 
to geometric and the algebraic averages. 
Namely, ${\langle \ln(x) \rangle  = \ln{\langle \langle x \rangle \rangle_{\tbox{g}}}}$  
and ${\mbox{Var}(x) = -2\ln(q)}$.
Given a hopping range ${|E_m-E_n| \le \omega}$ we can look for 
the typical matrix element $x_{\omega}$ for connected sequences 
of transitions, which we find by solving 
the equation ${ \varrho_{\tbox{E}} \omega \mbox{F}(x_\omega) \sim 1}$, 
where $\mbox{F}(x)$ is the probability to find 
a matrix element larger than~$x$. This gives 
\be{0}
x_{\omega} \ \ \approx \ \  
\langle \langle x \rangle\rangle_{\tbox{g}} 
\exp\left[2 \sqrt{-\ln q^{\alpha}} \right]
\ee
where $\alpha=\ln(\varrho_{\tbox{E}}\omega_c)$. 
From this equation we deduce the following: 
For $q{\lesssim}1$, meaning that the distribution 
is not too wide, $x_{\omega} \approx \langle \langle x \rangle\rangle_{\tbox{g}}$ 
as anticipated. But as the matrix gets more sparse (${q\ll1}$), 
the result deviates from the geometric average, 
the latter becoming merely a lower bound. 

The generalized VRH estimate is based 
on optimization of the integral 
$\int x_{\omega} \, \tilde{F}(\omega) \,d\omega$. 
For the rectangular $\tilde{F}(\omega)$ 
which has been assumed below Eq.(\ref{e223})
this optimization is trivial 
and gives $\approx x_{\omega_c}$, leading to
\be{12}
G_{\tbox{SLRT}} \ \ = \ \ 
q \ \exp\left[2 \sqrt{-\ln q^{\alpha}} \right]  
\times G_{\tbox{LRT}}
\ee
where $G_{\tbox{LRT}}$ is given by Eq.(\ref{e23}) 
and $q$ is given by Eq.(\ref{e24}).
We have also tested the standard VRH that assumes 
an exponential $\tilde{F}(\omega)$ (not presented).

\section{Numerical results}
\begin{figure}[h!t]
\centering

\includegraphics[clip,width=\hsize]{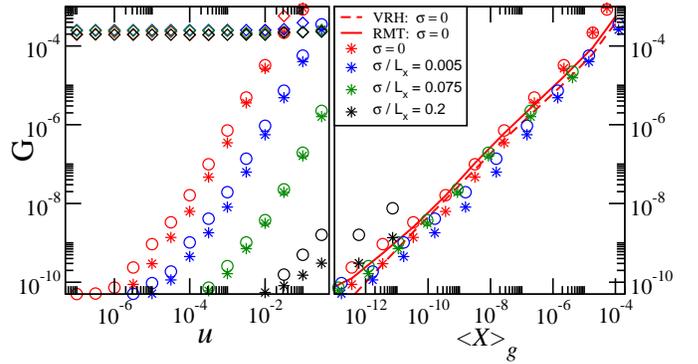}

\caption{
{\em Left panel:} 
The scaled $\tilde{G} \equiv \langle\langle x \rangle\rangle$ 
in the LRT and in the SLRT case as a function of $u$ for ${AS=1}$ 
and different smoothness of the deformation. 
The stars are for the physical matrices, while  
the circles are for their untextured versions (see text). 
The diamonds are for the LRT case.
{\em Right panel:} 
The SLRT result ${\langle\langle x \rangle\rangle}$ 
versus the geometric average ${\langle\langle x \rangle\rangle_{\tbox{g}}}$.  
These are compared with RMT based results, 
and with the associated analytical estimate of Eq.(\ref{e12}).    
We see the the agreement is very good. 
}

\label{fig:G_ratio1}
\end{figure}

\begin{figure}[h!t]
\centering
\includegraphics[clip,width=\hsize]{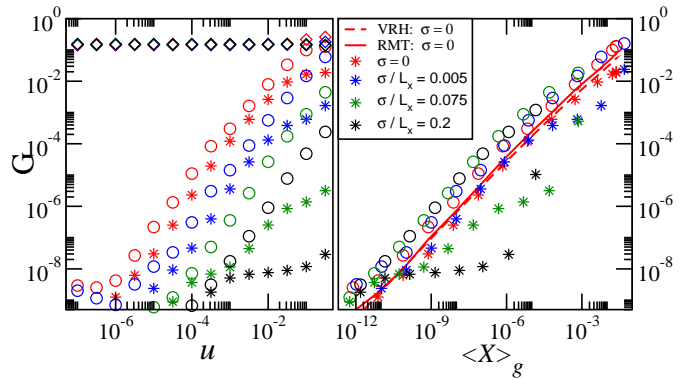}

\caption{
The same set of plots as in Fig.\ref{fig:G_ratio1} but for ${AS=20}$.
In the right panel we clearly see the departure of the physical 
result from the untextured and RMT results, 
and hence from the analytical estimate of Eq.(\ref{e12}). 
}

\label{fig:G_ratio20}
\end{figure}

The analytical estimates in Eqs.(\ref{e21},\ref{e22}) 
are supported by the histograms of Fig.\ref{fig:hist_delta_ratio_comp}.
For each choice of the parameters ${(AS, \sigma, u)}$ 
we calculate the {\em algebraic}, and the {\em geometric} 
and the {\em SLRT} resistor network averages of $\{|V_{nm}|^2\}$.
See  Fig.\ref{fig:G_ratio1} and Fig.\ref{fig:G_ratio20}. 
We also compare the actual results for~$G_{\tbox{SLRT}}$ 
with those that were obtained from a log-normal RMT ensembles 
with the same algebraic and geometric averages 
as that of the physical matrix \cite{d}.
As further discussed in the next paragraph one concludes 
that the agreement of the physical results
with the associated VRH estimate Eq.(\ref{e12}) is very good 
whenever the perturbation matrix is not textured, 
which is in fact the typical case for non-extreme 
aspect ratios.

In order to figure out whether the result is fully determined 
by the distribution of the elements or else 
texture is important we repeat the calculation 
for {\em untextured} versions of the {\em same}  matrices.  
The untextured version of a matrix is obtained by performing  
a random permutation of its elements along the diagonals. 
This procedure affects neither the bandprofile 
nor the $\{|V_{nm}|^2\}$ distribution, 
but merely removes the texture. 
In Fig.\ref{fig:G_ratio1} we see that the physical results 
cannot be distinguished from the untextured results, 
and hence are in agreement with the RMT and with the associated VRH estimate. 
On the other hand, in Fig.\ref{fig:G_ratio20}, 
which is for large aspect ratio, 
we see that the physical results deviate significantly 
from the untextured result. As the width of the Gaussian 
potential becomes larger (smoother deformation), 
the texture becomes more important. 
These observation are in complete agreement with the 
expectations that were discussed in the modeling section.

\section{Experiment}
As in \cite{nir1,nir2,nir3} a collection of $N{\sim}10^6$ atoms, 
say ${}^{85}Rb$ atoms (${\mass=1.4 \times 10^{-25} kg}$),
are laser cooled to low temperature of ${T \sim 10\mu K}$, 
such that the the typical thermal velocity is ${v_{\tbox{E}} \sim 0.05 m/s}$.
The atoms are trapped in an optical billiard whose blue-detuned light walls 
confine the atoms by repulsive optical dipole potential. 
The motion of the atoms is limited to the billiard plane by a strong perpendicular 
optical standing wave. The thickness of the billiard walls (${\sim 10 \mu m}$) 
is much smaller than its linear size (${L\sim 200\mu m}$). 
The 2D mean level spacing is ${\Delta = \varrho_{\tbox{E}}^{-1} \sim 2.5\times 10^{-34} J}$, 
which is ${2.4 Hz}$.
One or more of the billiard walls can be vibrated with several kHz frequency 
by modulating the laser intensity. The dimensionless spectral bandwidth 
of this driving can be set as say ${\omega_c/\Delta \sim 1000}$,  
with an amplitude ${\sim 10\mu m}$, such that ${\dot{R} \sim 0.015 m/s}$.
The temperature of the trapped atoms can then be measured as a function of time by 
the time-of-flight method. The LRT estimate ${G_{\tbox{LRT}} \sim 1.3 \times 10^{-51} Js/m^2}$ 
would lead to heating rate ${\dot{E} \sim 2 \times 10^{-27} J/s }$ 
which is ${\sim 0.15 mK/s}$. 
Considering (say) the geometry of Fig.\ref{fig:model}c,
the deformation ($u$) is achieved either by introducing an off center
optical ``spot", or by deforming slightly the optical walls (such precise control 
on the geometry has been demonstrated in previous experiments). 
Having control over $u$ we can have ${q \sim 10^{-5}}$ 
that would imply factor $100$ suppression, i.e. an estimated 
heating rate of few ${\mu K / sec}$. Such heating rate can be accurately measured, 
yielding high sensitivity to the energy diffusion process studied here.

\section{SLRT vs LRT} 
Typically the environment introduces in the dynamics an incoherent 
relaxation effect. If the relaxation rate is strong compared 
with the rate of the externally driven transitions, then the 
issue of having ``connected sequences of transitions"  becomes 
irrelevant, and the SLRT slowdown of the absorption is not expected.
In the latter case LRT rather than SLRT is applicable. 
It follows that for finite relaxation rate there is a crossover 
from LRT to SLRT behavior as a function of the intensity of the driving.
In cold atom experiments the relaxation effect can be controlled, 
and typically it is negligible. 
Hence SLRT rather than LRT behavior should be expected.  
This implies, as discussed above, a much smaller absorption rate. 
Furthermore, as discussed in the introduction, 
one can verify experimentally the signature of SLRT: 
namely, the effect of adding independent driving sources 
is expected to be non-linear with respect to their spectral content.

\section{Conclusions}
In this work we have introduced a theory for the calculation 
of the heating rate of cold atoms in vibrating traps. 
This theory, that treats the diffusion in energy space 
as a resistor network problem, is required if the cavity 
is not strongly chaotic and if the relaxation effect is small. 
The SLRT result, unlike the LRT (Kubo) result is extremely 
sensitive to the sparsity and the textures that characterize 
the perturbation matrix of the driving source.
For typical geometries the ratio between them 
is determined by the sparsity parameter~$q$ as in Eq.~(\ref{e12}), 
and hence is roughly proportional to the deformation ($u^2$) 
of the confining potential. 
If the cavity has a large aspect ratio, 
and the deformation of the confining potential is smooth,  
then the emerging textures in the perturbation matrix 
of the driving source become important, 
and then the actual SLRT result becomes even smaller.

By controlling the density of the trapped atoms, 
or their collisional cross section (e.g. via the Feshbach resonance) 
the atomic collision rate can be tuned by many orders of magnitude. 
Their effect on the dynamics can thus be made either negligible (as assumed above)  
or significant, thereby serving as an alternative (but formally similar) 
mechanism for {\em weak breakdown of integrability}.
It follows that heating rate experiments can be used 
not only to probe the deformation ($u$) of the confining potential, 
but also to probe the interactions between the atoms.


{\bf Acknowledgments. --} 
This research was supported by a grant from the USA-Israel
Binational Science Foundation (BSF).



\clearpage


\begin{thebibliography}{99}

\bibitem{wall1} 
J. Blocki, Y. Boneh, J.R. Nix, J. Randrup, M. Robel, A.J. Sierk and W.J. Swiatecki, 
Ann. Phys. {\bf 113}, 330 (1978). 

\bibitem{wall2} 
S.E. Koonin, R.L. Hatch and J. Randrup, 
Nucl. Phys. A {\bf 283}, 87 (1977).
%
S.E. Koonin and J. Randrup, 
Nucl. Phys. A {\bf 289}, 475 (1977). 

\bibitem{frc} 
D. Cohen, Annals of Physics 283, 175 (2000); cond-mat/9902168.


\bibitem{nir1} 
N. Friedman, A. Kaplan, D. Carasso, and N. Davidson, Phys. Rev. Lett. 86, 1518 (2001).

\bibitem{nir2}
A. Kaplan, N. Friedman, M. F. Andersen, and N. Davidson, Phys. Rev. Lett. 87, 274101 (2001).

\bibitem{nir3}
M. Andersen, A. Kaplan, T. Grunzweig and N. Davidson, Phys. Rev. Lett. 97, 104102 (2006).

\bibitem{wls} 
A. Barnett, D. Cohen and E.J. Heller, 
Phys. Rev. Lett. {\bf 85}, 1412 (2000); J. Phys. A {\bf 34}, 413 (2001).


\bibitem{drifts}
T. A. Savard, L. M. Ohara and J. E. Thomas,
Phys. Rev. A {\bf 56}, R1095 (1997).


\bibitem{shake} 
S. Friebel, C. D'Andrea, J. Walz, M. Weitz, and T. W. Hansch, 
Phys. Rev. A {\bf 57}, R20 (1998).


\bibitem{louisell}
W.H. Louisell, 
{\it Quantum Statistical Properties of Radiation}, 
(Wiley, London, 1973).


\bibitem{slr1} 
D. Cohen, T. Kottos and H. Schanz, J. Phys. A 39, 11755 (2006).
M. Wilkinson, B. Mehlig, D. Cohen, Europhysics Letters 75, 709 (2006).

\bibitem{slr2}
S. Bandopadhyay, Y. Etzioni and D. Cohen, Europhysics Letters 76, 739 (2006).
A. Stotland, R. Budoyo, T. Peer, T. Kottos and D. Cohen, J. Phys. A 41, 262001 (FTC) (2008).


\bibitem{miller}
A. Miller and E. Abrahams, Phys. Rev. {\bf 120}, 745 (1960).

\bibitem{mott}
N.F. Mott, Phil. Mag. {\bf 22}, 7 (1970). 

\bibitem{ambeg}
V. Ambegaokar, B. Halperin, J.S. Langer, 
Phys. Rev. B {\bf 4}, 2612 (1971). 
%
M. Pollak, J. Non-Cryst. Solids {\bf 11}, 1 (1972).



\vspace*{1mm}



\bibitem[a]{a}
Our interest is in systems that are classically chaotic. 
This  means exponential sensitivity to change in initial conditions, 
without having mixed phase space. 


\bibitem[b]{bb}
For a more general version of ${\dot{E}=D/T}$,
that does not assume a Boltzmann-like distribution
with a well defined temperature, 
see section IV of Ref.\cite{frc}.
   

\bibitem[c]{b}
The average is taken over all the elements within the 
energy window of interest as determined by the preparation temperature.   
The weight of $|V_{nm}|^2$ in this average is determined 
by the spectral function as ${\tilde{S}(E_n{-}E_m)}$. 


\hide{
\bibitem[d]{c}
In the original application of SLRT, $R$ is the magnetic 
flux through a ring, $\dot{R}$ is the electromotive force, 
and $G$ is the mesoscopic conductance (up to a factor).  
}


\bibitem[d]{cc}
For a very small $u$, an optional route 
that bypass the resistor network calculation,  
is to analyze the slow ($\propto u^2$) transitions between 
noise-broadened energy levels.

\bibitem[e]{d}
Since for the log-normal distribution the median equals the geometric
average, we used the median in the definition of $q$ for the sake of the
numerical stability.



\end{thebibliography}
\end{document}